\documentclass[pra,twocolumn,aps,superscriptaddress]{revtex4}
\usepackage{graphicx}
\usepackage{amsmath}
\usepackage{amssymb}
\usepackage{hyperref}\hypersetup{colorlinks=true,linkcolor=blue,citecolor=blue,pdfstartview=FitH}
\usepackage{bm}\renewcommand{\mathbf}{\bm}
\usepackage{subfigure}

\newcommand{\figwidth}{0.95\columnwidth}
\newcommand{\largefigwidth}{0.85\columnwidth}

\DeclareMathOperator{\Tr}{Tr}

\DeclareMathOperator{\Real}{Re}\renewcommand{\Re}{\Real}
\DeclareMathOperator{\Imag}{Im}\renewcommand{\Im}{\Imag}

\newcommand{\be}{\begin{equation}} \newcommand{\ee}{\end{equation}}
\newcommand{\bea}{\begin{eqnarray}} \newcommand{\eea}{\end{eqnarray}}

\begin{document}
\title{Prethermalization and glassiness in the bosonic Hubbard model}
\author{I. Salazar Landea\footnote{peznacho@gmail.com}}
\author{N. Nessi\footnote{nnessi@fisica.unlp.edu.ar}}
\affiliation{Instituto de F\'{\i}sica La Plata (IFLP) - CONICET and Departamento de F\'{\i}sica,\\
Universidad Nacional de La Plata, CC 67, 1900 La Plata, Argentina.}

\begin{abstract}
We investigate the non-equilibrium dynamics of the bosonic Hubbard model starting from inhomogeneous superfluid or Mott insulator initial states using the truncated Wigner approximation (TWA). We find that the relaxation of the system develops in two steps for sufficiently large interaction strengths: after a fast relaxation the system gets caught in metastable prethermalized states that precede the true equilibrium state. We find that the lifetime of these prethermalized states increases by several orders of magnitude as we increase the on-site interaction strength beyond a threshold value. We show that the emergence of long-lived metastable states in the quantum dynamics is associated with an ergodic (active) to non-ergodic (inactive) dynamical phase transition in the ensemble of classical trajectories that contribute to the semiclassical limit. This dynamic phase transition, which is very similar to that found in different classical models of glasses, is closely related to the dynamic heterogeneity of the classical relaxation.
\end{abstract}


\maketitle

\section{Introduction}

The non-equilibrium dynamics of closed quantum systems is nowadays a very active field of research. The motivations are manifold. On one hand, the determination of the necessary conditions for a closed quantum systems to reach asymptotically a state of thermal equilibrium constitutes a key issue in the study of the foundations of statistical mechanics. On the other hand, with the advent of cold atomic systems the manipulation of nearly perfectly isolated quantum systems with large coherence times has become experimentally feasible. There are several theoretical and experimental examples of systems that starting from non-equilibrium conditions reach, on accessible timescales, a state that is compatible with thermal equilibrium~\cite{trotzky11_relaxation_BH_10101010,sorg14_thermalization_bose-hub,rigol08_mechanism_thermalization,rigol14_quenches_thermo_limit}, at least at the level of simple correlation functions. However, situations in which thermal equilibration is elusive are clearly of particular interest. For integrable systems, thermal equilibration is not expected since the large number of constants of motion prevent the system from erasing memory of the initial conditions. However, there are examples of systems that are not integrable (but are close to an integrable point) for which thermal equilibration is not observed on the accessible timescales. Maybe the first experimental example of a (non-integrable) closed quantum many-body system failing to reach thermal equilibrium on accessible timescales was provided by the famous experiment of Kinoshita et al.~\cite{kinoshita06_non_thermalization}.

The phenomenon of prethermalization~\cite{berges04_prethermalization_idea,moeckel08_quench_hubbard_high_d,eckstein09_thermalization_quench_fermi_hubbard,kollar11_gge_pretherm,nessi14_gge_fermi_liquid,gring12_prethermalization_isolated_bose_gas,essler13_quench_tuneable_integrability_breaking,marcuzzi13_pretherm_ising} is one of the possible explanations for the apparent lack of thermalization in certain nearly integrable systems. Indeed, in certain situations it can be shown that the relaxation of systems close to an integrable point may proceed in two steps. The first step is a fast relaxation whose main mechanism is dephasing between some quasifree modes of the system~\cite{nessi14_gge_fermi_liquid}. Close to an integrable point, where other relaxation mechanism such as inelastic collisions are extremely inefficient, such dephasing relaxation often gives rise to a highly non-thermal metastable state. Afterwards, the remaining relaxation channels become relevant and the prethermal metastable state starts to decay to the final state of the system~\cite{moeckel08_quench_hubbard_high_d,eckstein09_thermalization_quench_fermi_hubbard,stark13_kinetic_description}. One of the most characteristic phenomenons associated with prethermalization is the fact that certain observables reach its final, thermal value in the short timescales of the dephasing relaxation. The kinetic energy of the system is the most prominent example of such behavior. Prethermalization has been observed in experiments with ultracold bosonic gases~\cite{gring12_prethermalization_isolated_bose_gas,langen15_quench_lieb-liniger}.

In particular, prethermalization may be behind the apparent breakdown of thermalization in the strongly interacting Bose-Hubbard model, which has received special attention. Starting with the pioneering work of Kollath et al.~\cite{kollath07_quench_BH}, where using density matrix renormalization group (DMRG) techniques it was shown that for sufficiently strong interactions few-body correlation functions reached a quasistationary state that was not compatible with thermal equilibrium, several studies have focused on this system~\cite{roux09_quench_bose-hub,carleo2012localization}.  In Ref.~\cite{carleo2012localization} it was shown that for strong interactions the system gets trapped in long-lived inhomogeneous metastable states. Such kinetic arrest, was shown~\cite{carleo2012localization} to be associated with a dynamical localization in the many-body Hilbert space. The Bose-Hubbard model is also one of the most extensively realized models in cold atoms experiments~\cite{bloch08_cold_atoms_optical_lattices_review}. In particular, experiments with inhomogeneous Bose-Einstein condensates also demonstrate pronounced timescale separation and slow thermalization for strong interactions~\cite{natu11_local_eq_inhom_BEC,hung10_slow_transport_inhom_BEC,mckay15_metastability_inhom_BEC}.

On the other hand, since the seminal work of Srednicki about the eigenstate thermalization hypothesis~\cite{srednicki94_ETH}, semiclassical considerations have been very important in our understanding of quantum thermalization. Recently Cosme et. al~\cite{cosme2014thermalization} made a contribution in this line through the truncated Wigner approximation (TWA) applied to a two sites multi-band bosonic Hubbard model. Since the TWA allows to consider the first order quantum corrections to the classical action~\cite{polkovnikov2010phase} the connection between quantum physics and its classical limit becomes explicit. In particular, it has been checked in~\cite{cosme2014thermalization} that the thermalization of the quantum system comes along the ergodicity of classical trajectories.
Also along these lines, of particular relevance to the study presented here are the works~\cite{jona1996chaotic} and~\cite{cassidy09_quench_bose-hub}  where the breakdown of thermalization for strong interactions was linked with the chaotic properties of the mean-field limit. More precisely, it was shown that the Lyapunov exponent of the classical trajectories associated with the solutions of the mean-field equations is suppressed for strong interactions.

The appearance of long-lived prethermalized states in the relaxation of isolated quantum systems is phenomenologically very similar to the physics of classical glassy systems such as spin glasses~\cite{castellani05_spin-glass_pedestrians}, atomistic glass formers~\cite{cavagna09_supercooled_pedestrians,chandler10_dyn_pt_review} and kinetically constrained models~\cite{berthier11_dyn_heterogeneity_kcm}. Such systems exhibits a typical two-step relaxation: after a fast inertial regime (analogous to the dephasing regime) they get caught in metastable states whose lifetime can be astronomical in certain parameter regions, for example, for sufficiently low temperature or high density in the case of atomic models. One of the most distinctive characteristics of glassy relaxation is the appearance of dynamical heterogeneity~\cite{berthier11_dyn_heterogeneity_kcm}, i.e., the fact that the relaxation is spatially non-homogeneous: fast and slow regions are clustered together. One of the most appealing theories to explain such feature is that of a \emph{dynamical} phase transition between active and inactive phases taking place in the ensemble of possible histories that the system can follow in the relaxation~\cite{garrahan07_dyn_pt_kcm,garrahan09_dyn_pt_long,hedges2009dynamic,chandler10_dyn_pt_review}.

In this work we revisit the problem of the apparent breakdown of thermalization in the bosonic Hubbard model from the semiclassical perspective provided by the TWA. For large occupation numbers, this approach allows to access the dynamics of the quantum system performing weighted averages over ensembles of classical (mean field) trajectories. We explicitly show that prethermalized states can have a huge lifetime that grows \emph{exponentially} while increasing the coupling strength, which prevents the observation of thermal equilibration on any reasonable timescale. Moreover, we find that the emergence of long-lived prethermalized states is closely related with the appearance of non-ergodic, glassy features in the mean-field trajectories associated with the quantum dynamics. More specifically, we find that the ensemble of mean-field trajectories undergo a dynamical phase transition between an active and an inactive phase, completely analogous to that observed in classical models of glasses. We find that in the present case this transition is also intimately correlated with a remarkable phenomenon present in the mean field trajectories, dynamical heterogeneity. We shall go deeper into these issues in the remaining part of the paper, which is organized as follows: In Section~\ref{sec:quantum} we describe the model, the initial conditions, the semiclassical approximation used to study the quantum dynamics and show the results for the quantum dynamics, while in Section~\ref{sec:classical} we thoroughly investigate the properties of the ensemble of mean-field trajectories that contribute to the quantum dynamics and show that they show clear signatures of glassiness as we increase the strength of the interaction. In Section \ref{conc} we briefly sumarize our results.

\section{Quantum dynamics}\label{sec:quantum}

%\subsection{Truncated Wigner Approximation}

Let us consider the $1D$ bosonic Hubbard model with Hamiltonian $H=H_0+H_{int}$:
\bea
\label{qh}
\nonumber H_0&=&-J\sum_{j=1}^{L}\left(a_j^\dagger a_{j+1}+a_{j+1}^\dagger a_{j} \right),\\
H_{int}&=&U \sum_{j=1}^{L}n_j n_j\,,
\eea
where $L$ is the number of sites in the chain, the $a$'s are canonical bosonic operators and $n_j = a_{j}^\dagger a_{j}$. We shall consider periodic boundary conditions, $a^{\#}_{L+m}=a^{\#}_m$. We shall analyze the dynamics generated by~\eqref{qh} starting from an inhomogeneous initial state described by a density matrix $\rho_0$, which we left unspecified until the next section. In particular, we shall focus on the relaxation of the density profile $\langle n_j(t)\rangle=\Tr[\rho_0 e^{iHt}n_je^{-iHt}]$, where we have set $\hbar=1$. The thermal density profile is given by $\langle n_j\rangle_{\mathrm{th}}=N_a/L=N$ for all $j$, where $N_a$ is the total number of bosons.

When the average number of atoms per site $N$ is large the dynamics of the Bose-Hubbard model can be studied using the TWA \cite{polkovnikov2002evolution}. To calculate the time dependence of expectation values within the TWA we have to consider the solution of the classical equations of motion associated with the quantum Hamiltonian~\eqref{qh}. These are the standard lattice Gross-Pitaevskii equations \cite{polkovnikov2002evolution},
\bea
\label{eom}
i\frac{d\psi_j(t)}{dt}=-J(\psi_{j+1}(t)+\psi_{j-1}(t))+2 U \psi_j(t)\left| \psi_j(t) \right|^2\,.
\eea
where the classical fields are normalized to the total number of particles $\sum_{j=1}^L\left| \psi_j(t)\right|^2=N_a$. Then the expectation value of any given operator $\Omega$ at time $t$ can be calculated averaging the corresponding classical observable $\Omega_{cl}$ over an ensemble of initial conditions weighted according to the Wigner transform of the initial density matrix $\rho_0$
\bea
\Omega(t)=\int d\psi_0^* d\psi_0\, p\left(\psi_0,\psi_0^*\right)\Omega_{cl} \left( \psi(t), \psi^*(t) \right)\,,
\eea
where $p$ is defined as
\bea\label{twa}
\nonumber
p\left(\psi_0,\psi_0^*\right)&=&\int d\eta_0^* d\eta_0 \left< \psi_0-\frac{\eta_0}{2}\right|\rho_0 \left|\psi_0+\frac{\eta_0}{2}\right>\\
&&\times\, e^{-\left|\psi_0\right|^2-\frac14 \left|\eta_0\right|^2}e^{\frac12\left(\eta_0^*\psi_0-\eta_0\psi_0^*\right)}\,.
\eea
To lighten the notation we have omitted the site index $j$. The measure is
\bea
 d\psi_0^* d\psi_0= \prod_{j=1}^L d\psi_{j}^*(0) d\psi_{j}(0) \,.
\eea

The correspondence between classical and quantum observables can be formulated most easily using the coherent state Bopp representation, which makes the assignments
\bea
\hat{a}^\dagger&\rightarrow&\psi^*-\frac12 \frac{\partial}{\partial \psi}\,,\\
\hat{a}&\rightarrow&\psi+\frac12 \frac{\partial}{\partial \psi^*}\,,
\eea
so that
\bea
n_j\rightarrow |\psi_j|^2-\frac12\,.
\eea

The TWA is the leading order approximation in an expansion in a small parameter that measures deviations from classicality \cite{polkovnikov2010phase}. In the particular case of interacting bosons this parameter is $1/N$, the inverse of average number of bosons per site. The TWA is a controlled approximation in the sense that it is possible to calculate higher order corrections that take the form of stochastic perturbations to the classical trajectories \cite{polkovnikov2010phase}. However, as any approximation, be it controlled or not, there are certain parameter regimes in which it works better than others. To define when we expect this approximation to be accurate we may introduce the nonlinearity parameter $\lambda=\frac{UN}{J}$ that is the ratio between the typical potential energy per site and the typical kinetic energy per site~\cite{cassidy09_quench_bose-hub}. When the interactions are strong enough, $\lambda\sim N^2$, the system (in equilibrium) undergoes a quantum phase transition to a Mott insulating state. In the vicinity of the transition quantum fluctuations become large and we cannot expect the TWA to work well in this case, i.e., second order corrections become important in that regime. However, there is a wide regime $\lambda\lesssim N^2$ ($U/J\lesssim N$) where the ground state of the system is a superfluid (weakly or strongly interacting depending on whether $\lambda<1$ or $\lambda>1$ respectively) where we can expect the TWA to be a good approximation to the dynamics. We shall work in that regime.

In parallel, the choice of the initial condition is also relevant for the accuracy of the results. We shall work with two types of initial states, coherent (superfluid) initial states and Fock (Mott insulator) initial states. The coherent state represents the initial condition that is closest to a perfectly defined classical initial condition and therefore we can expect the TWA to accurately capture the dynamics. In particular, the Wigner function is
\bea
p_C(\psi_0,\psi_0^*)=\prod_{j=1}^L \frac{2}{\pi}e^{-2\left| \psi_j-N_j \right|^2}\,,
\eea
where $N_j=\langle n_j(0)\rangle$. The Wigner function is a true probability distribution and can be sampled efficiently as $\psi_j=\sqrt{N_j}+\frac{1}{2}(\eta_1+i\eta_2)$~\cite{cosme2014thermalization,olsen09_wigner_sampling} where $\eta_1$ and $\eta_2$ are two real Gaussian random variables.

\begin{figure}[htp]
\begin{center}
\includegraphics[width=\figwidth]{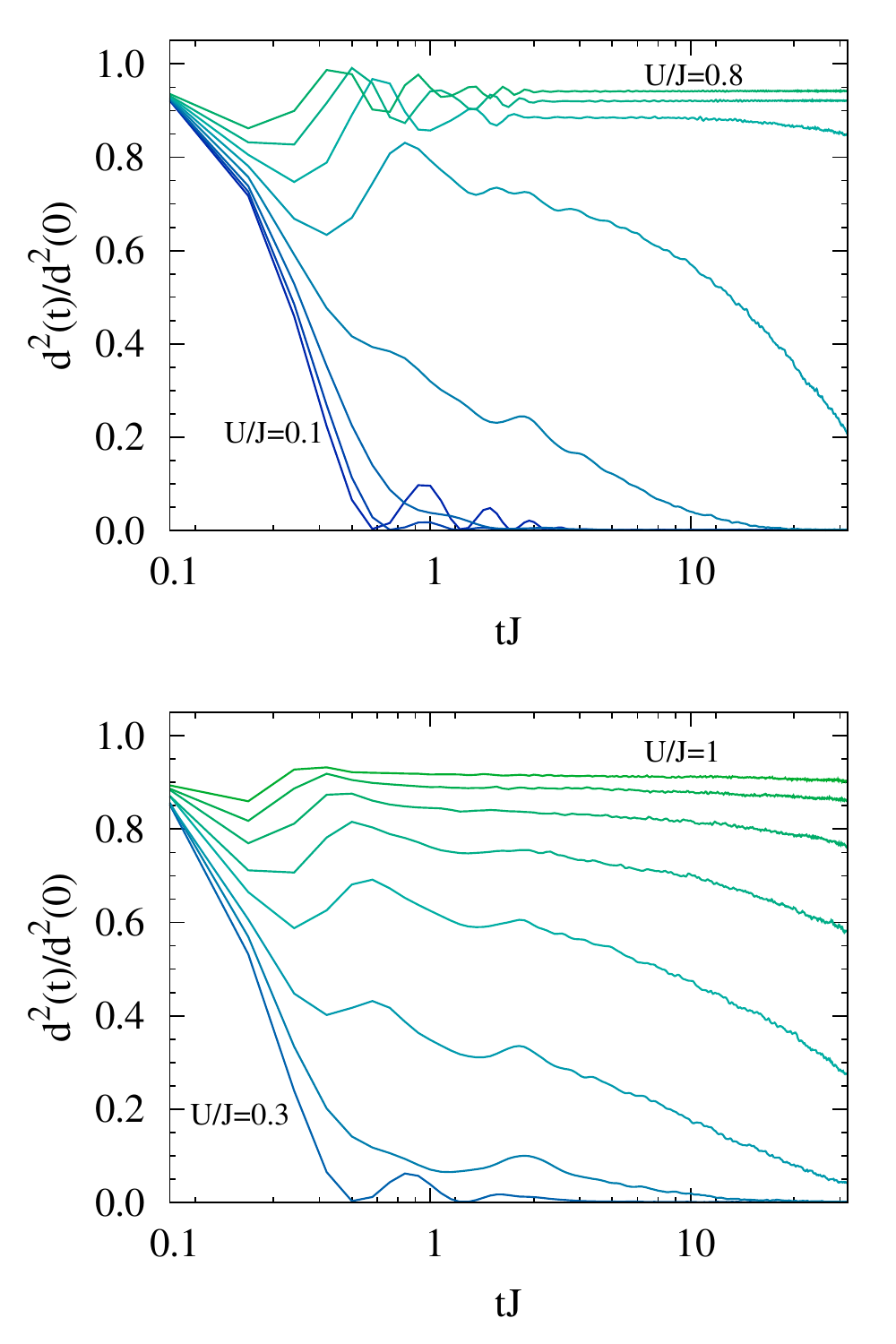}
\caption{\label{dvst}Dynamical distance $d(t)$ as a function of time
for the Fock (up) and coherent (down) initial states. Different curves
correspond to different interaction strengths $U/J$.}
\end{center}
\end{figure}

The Fock initial state is the least classical initial state in the sense that the initial phases are completely unspecified. The Wigner function in this case is
\bea
p_F(\psi_0,\psi_0^*)=\prod_{j=1}^L 2e^{-2\left| \psi_j \right|^2}L_{N_j}\left(4\left| \psi_j \right|^2\right)\,,
\eea
where $L_N(x)$ is the $N^{th}$ order Laguerre polynomial. Unfortunately, this $p$ generically is not definite positive and for large $N$ has a highly oscillatory behavior. This fact makes convergence much slower when doing a Monte Carlo integration over initial conditions. To bypass this issue we shall approximate this Wigner function by products of true distributions reproducing the first moments of the Laguerre polynomials. The simplest way to do this is just to reproduce the first moment by replacing each factor $p_{N_j}(\psi_j^*,\psi_j)$ by a Dirac delta function $\delta\left(|\psi_j|^2-\frac12-N_j\right)$ and averaging over random phases when doing the Monte Carlo integration~\cite{polkovnikov2003evolution,cosme2014thermalization,olsen09_wigner_sampling}. The next step would be considering a distribution with two moments, i.e., the Gaussian $\frac{2}{\sqrt{2\pi}}e^{-2\left(|\psi_j|^2-\frac12-N_j\right)^2}$. We will refer to this two approximations as the delta and Gaussian Wigner functions $p_\delta$ and $p_g$ respectively.
We have checked that in the particular cases that we investigated both approaches give very similar results. We shall therefore use the delta function approximation due to computational convenience. For the Fock initial state it was found that, in the worst case, the TWA results quantitatively deviate from available exact solutions for times larger than $t_c=J/U$~\cite{polkovnikov2002evolution}. However, the predictions of the TWA remain qualitatively valid for larger times.

We shall now proceed to investigate the quantum dynamics of the density profile. We shall chose the coherent or Fock initial states in such a way that  the initial density profile is $(20, 0,20,0,\dots)$ : an alternation of empty and highly occupied sites. With this election $N=10$. The system is on a ring of size $L=30$. This initial state may be experimentally relevant for cold atomic gases loaded on optical lattices~\cite{trotzky11_relaxation_BH_10101010}. To calculate the quantum dynamics we sample $10^3$ different initial conditions both for the coherent and for the Fock initial state, but in order to test convergence we went up to $10^4$  realizations.
To quantify the overall relaxation of the density profile to its thermal configuration we introduce the dynamical distance:
\bea
d^2(t)=\frac{1}{L}\sum_{j=1}^L \left(n_j(t)-N\right)^2\,.
\eea
Clearly, thermal equilibrium implies $d^2(t)=0$. In Fig.~\ref{dvst} we show the decay of $d^2(t)$ for both the coherent and the Fock initial states for different values of the coupling strength. In both cases for low coupling strength the system thermalizes quickly in a timescale of the order of one hopping~\cite{trotzky11_relaxation_BH_10101010,sorg14_thermalization_bose-hub}. As we increase the coupling strength a metastable state emerges in between the initial relaxation and the final decay to equilibrium. The lifetime of such metastable state increases with the coupling strength until it becomes larger than the maximum time available in our simulation. For the Fock initial state the increase of the lifetime of the metastable state is more abrupt. In both cases, the density profile of the metastable state is closer to that of the initial state as we increase the coupling strength.

\begin{figure}[htp]
\begin{center}
\includegraphics[width=\largefigwidth]{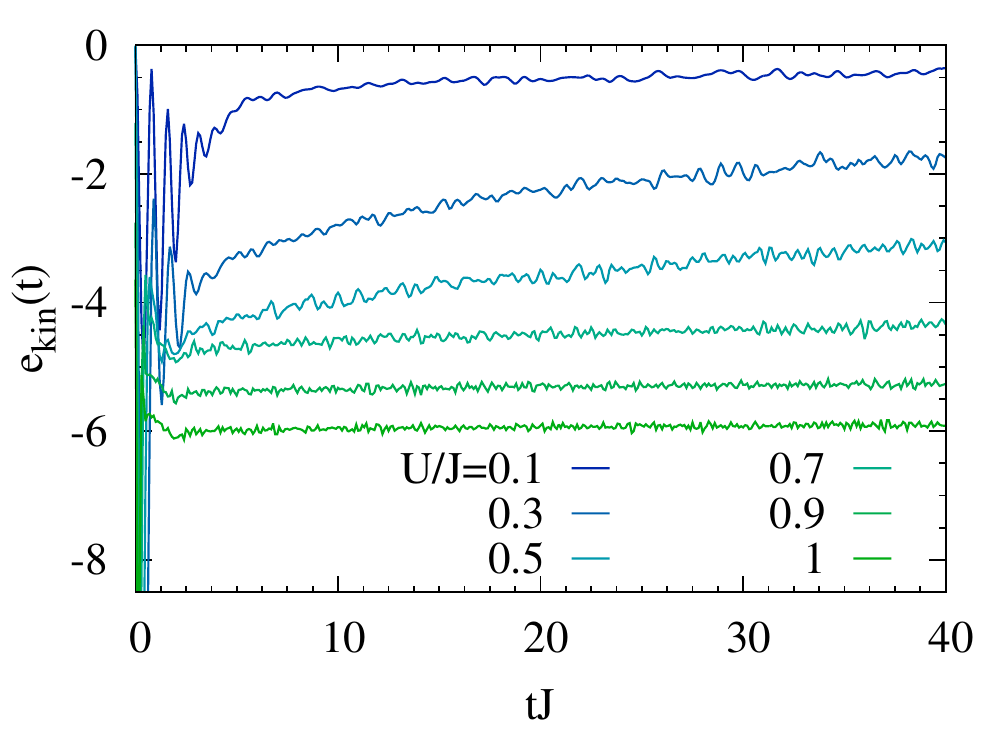}
\caption{\label{ekin}Relaxation of the kinetic energy for the coherent state initial condition. Different curves
correspond to different interaction strengths $U/J$. Curves are vertically shifted for clarity.}
\end{center}
\end{figure}

\begin{figure}[htp]
\begin{center}
\includegraphics[width=\figwidth]{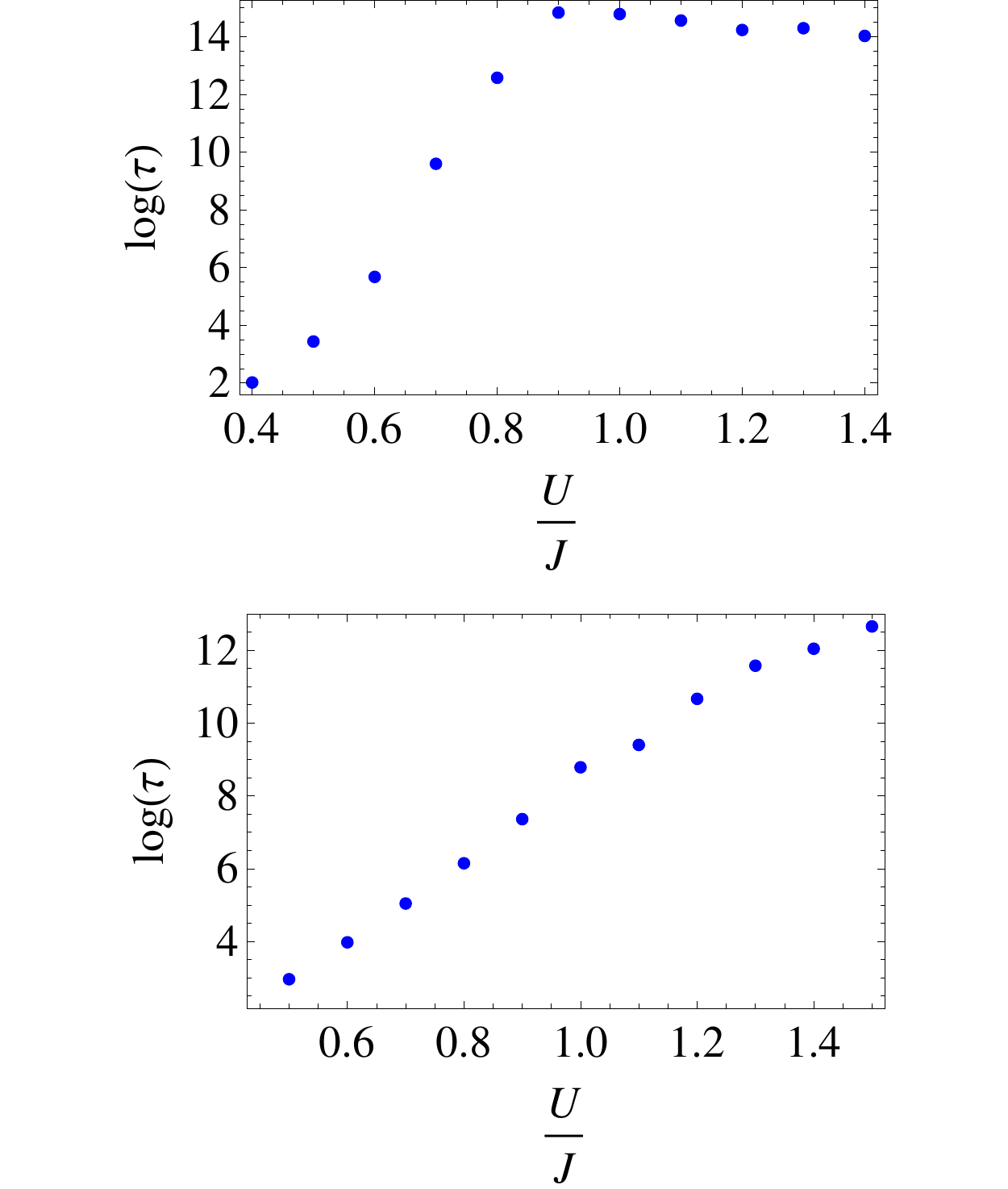}
\caption{\label{tauro}Logarithm of the relaxation time as a function
of coupling $U/J$ for the Fock (up) and coherent (down) initial
state. }
\end{center}
\end{figure}

In Fig.~\ref{ekin} we show the relaxation of the kinetic energy $e_{kin}(t)=\Tr[\rho_0 e^{iHt}H_0e^{-iHt}]$ for the coherent state initial condition for several values of the interaction strength. For the Fock initial state we obtain qualitatively the same results. We observe that for weak interactions $U/J\lesssim 0.1$ the kinetic energy prethermalizes in a timescale $t^{weak}_{pt}\sim 10 J^{-1}$. For intermediate interaction strengths $0.2\lesssim U/J\lesssim 0.7$ the kinetic energy also tends to a well defined constant but with a larger relaxation time of the order of $t^{int}_{pt}\sim 30J^{-1}$ ($int$ stands for intermediate). Whereas for strong interaction strengths $U/J\gtrsim0.7$ the prethermalization timescale is $t^{strong}_{pt}\sim 3J^{-1}$. The variety of prethermalization timescales can be understood from the fact that the quasifree modes behind the dephasing relaxation are qualitatively different for strong and weak interactions. For weak interactions the quasifree modes are related to the momentum eigenmodes that diagonalize $H_0$ and are completely delocalized in space. For strong interactions, the quasifree modes are related with the eigenmodes that diagonalize $H_{int}$ and are completely localized excitations. For intermediate interactions there is a truly non-trivial regime for which the quasifree modes are neither completely localized nor delocalized. This suggests that the effective models that describe the short time dynamics in the three cases are completely different. However, a remarkable fact is that for strong interactions the prethermalization time is the same regardless the specific value of the coupling strength, while the relaxation timescale of the density profile shows large variations (see Fig.~\ref{tauro}) due to the presence of metastable states.

The crossover from fast thermalization to the metastable state
dominated regime for the two different initial conditions occur in
different coupling ranges. To obtain a decay timescale of $d^2(t)$
we fit  an exponential to the tail of the decay.
 In Fig.~\ref{tauro} we show the results for the decay timescales for both initial conditions. We find that the decay timescales increase by several order of magnitude as we increase the coupling strength in a small range.
%In order to be able to fit the relaxation time up to such long scales we integrated the equations of motion up to large enough times so that we could distinguish the average decay from the fluctuations.% with characteristic time scale $J\,\Delta t$.
%For large $U/J$ it was necessary to integrate the equations of motion up to  $J\, t_{obs}=320$. The inset in Figure \ref{tauro} shows an example of the fit for one of the most extreme cases.
We find that the center of the crossover region is around $U/J\simeq 0.7$ for the Fock initial state and $U/J\simeq 1$ for the coherent state. The emergence of prethermalized states with such large lifetime may completely hinder the observation of the thermal equilibrium state of the system after the quench and certainly is at the root of the apparent lack of thermalization observed in earlier works~\cite{kollath07_quench_BH,carleo12_glassy_bose-hubbard}. We find the situation rather similar to that of glasses, systems that exhibit the typical two-step relaxation as a consequence of getting caught in extremely long-lived metastable states, whose lifetime can be astronomical for sufficiently low temperatures or high densities. In the next section we shall outline a more qualitative relation between the emergence of long-lived metastable states and glassiness by analyzing the properties of the mean field trajectories associated with the quantum dynamics.

%the distance develops a plateau for large enough values of $U$. This intermediate time plateau is well described by an exponential $e^{-\frac{t}{\tau}}$, so that we can define the relaxation time $\tau$.
%The dependence of $\tau$ as a function of $U$  is shown in Figure \ref{tauro}.

%Let us make a few remarks on the regimes of times and interaction where there are no plateaus. Firstly, being the system not integrable we expect that it will always thermalize for large enough times, so that the distance will oscillate around zero for large times as we can see for small $U$ in Figure \ref{dvst}. On the other hand, for small enough $U$ we see that the system thermalizes before developing any plateau.

\section{Glassy properties in the mean-field trajectories}\label{sec:classical}

In Ref.~\cite{cosme2014thermalization} it was proposed that the ability of a quantum system to thermalize is related with the ergodicity of the classical trajectories of the associated mean-field system, an idea that is also implicit in previous studies, such as~\cite{jona1996chaotic} and~\cite{cassidy09_quench_bose-hub}. In this Section we shall
show that the emergence of long-lived prethermalized states in the quantum dynamics of the Hubbard model is correlated with the
lack of ergodicity of the classical trajectories that are used to
build up the quantum dynamics.

\begin{figure}[htp]
\begin{center}
\includegraphics[width=\figwidth]{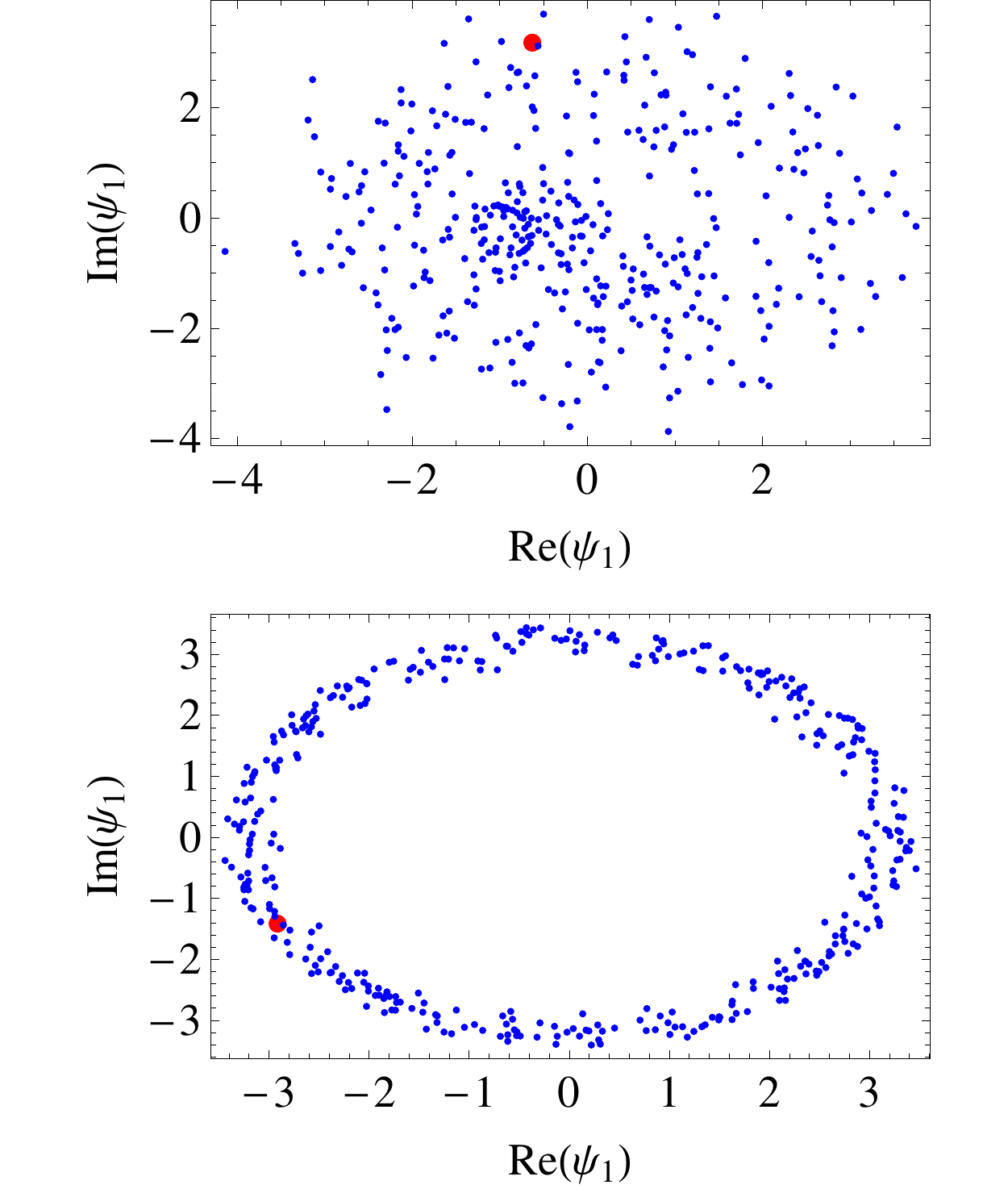}
\caption{\label{ps} Position in the phase space $\Re(\psi_1)$ vs.
$\Im(\psi_1)$ for the first site of the chain observed every a
$J\,\Delta t =0.1$ for $J\,t_{obs}=40$ for $U/J=0.5$ (up) and $U/J=1.5$
(down). The big red dot corresponds to the initial configuration.
Each site was chosen to have a random phase as initial
condition.}\label{fig:phase_space}
\end{center}
\end{figure}

To qualitatively introduce the discussion, in Fig.~\ref{ps} we show a stroboscopic sampling of the position in phase space of the field $\psi_1$ for two different couplings. We can see that for small coupling ($U=0.5$) the system explores the whole phase space while for large coupling  ($U=1.5$) it remains captured in a ring of finite width in the  $\Re(\psi_1)$ vs. $\Im(\psi_1)$  space. A similar behavior is observed when analyzing the behavior of the phase space trajectory of the other sites. Going back to Fig.~\ref{dvst} we can check that indeed for  $U=0.5$ the quantum density profile is thermalized for $tJ=40$ while it is still trapped in the metastable state for $U=1.5$.

Additionally, we point out that the phase of the fields $\theta_j(t)=\arctan(\Im[\psi_j(t)]/\Re[\psi_j(t)])$ turns out to be always ergodic, in the sense that along the dynamic evolution it visits all values between $0$ and $2\pi$, irrespective of the value of the coupling strength, for all trajectories. This fact is also illustrated by Fig.~\ref{fig:phase_space} where, even for the strong coupling regime, the phase of the field $\psi_1$ explores all values. The lack of ergodicity is thus encoded in the dynamics of the particle density at each site $n_j(t)=\vert \psi_j(t)\vert^2$.

We will now proceed to make a more detailed and quantitative analysis of the properties of the classical trajectories. In order to characterize them we introduce the mobility $K$ of each trajectory
\bea
K\left[\psi(t)\right]=\Delta t \sum_{t=0}^{t_{obs}} \sum_{j=1}^N \left(|\psi_i(t+\Delta t)|^2-(|\psi_i(t)|^2 \right)^2\,,
\eea
where $t_{obs}$ is the maximum observation time, i.e., the extension of the trajectories and $\Delta t$ is a UV cutoff that kills possible small oscillations leaving just the actual displacement in the phase space. % We set $\Delta t J=1$.
$K\left[\psi(t)\right]$ is an extensive-in-time quantity that
will typically be large for ergodic trajectories, while it will be generically small for non-ergodic trajectories as can be intuitively seen from Fig. \ref{ps}. Thus, $K\left[\psi(t)\right]$ can be used as an order parameter to distinguish ergodic (mobile) from non-ergodic (immobile) trajectories.

Since we are interested in assessing the properties of the trajectories that are most relevant to the quantum motion we will introduce a measure on the space of classical trajectories given by the Wigner function of the quantum initial state that we wish to consider. In the case of the coherent state, the measure $p_C(\psi_0,\psi_0^*)$ is well defined while in the case of the Mott insulating state we can choose any well defined distribution that approximates $p_F(\psi_0,\psi_0^*)$, such as $p_\delta$ or $p_g$. For simplicity, we shall restrict in the following analysis to the coherent state case.

\begin{figure}[htp]
\begin{center}
\includegraphics[width=\largefigwidth]{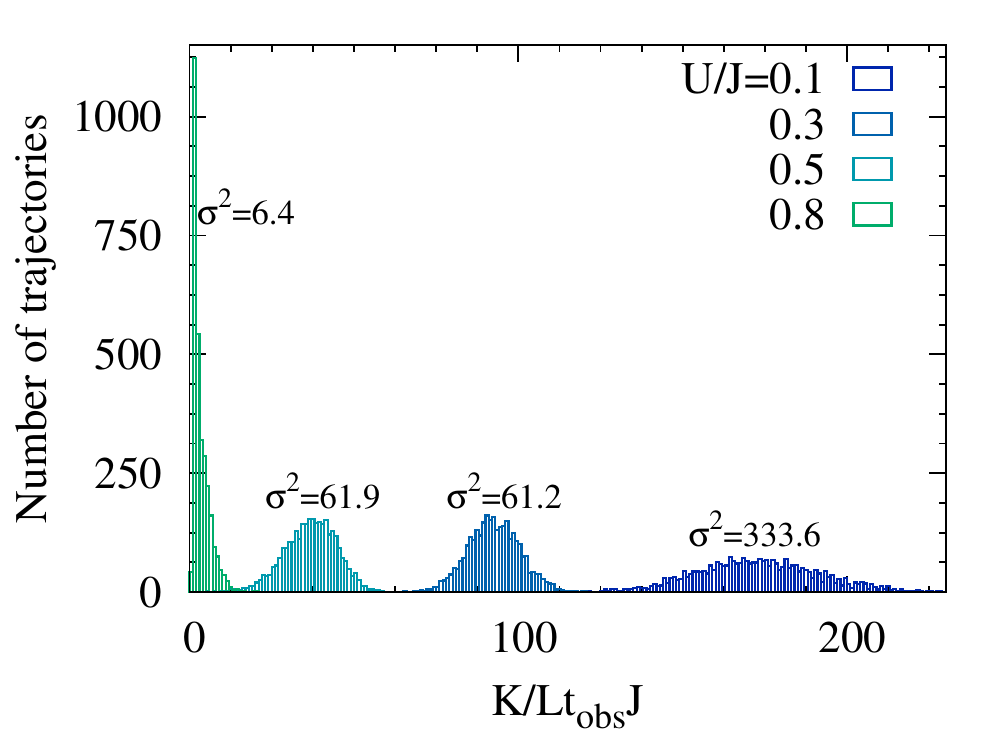}
\caption{\label{ps} Histogram of the mobility $K[\psi(t)]$ for ensembles of $2\times10^3$ trajectories sampled according to the Wigner function of the coherent state initial condition $p_C(\psi_0,\psi_0^*)$ for four different values of $U/J$.}\label{fig:histo}
\end{center}
\end{figure}

In Fig.~\ref{fig:histo} we show the mobility histogram from an ensemble of $2\times10^3$ trajectories sampled according to $p_C(\psi_0,\psi_0^*)$ for four different values of $U/J$. The center of the distributions monotonically shifts to lower values of the mobility as we increase $U/J$, but the dispersion around the center decreases for large coupling constants. In particular, for sufficiently large coupling all the trajectories have a very small mobility with a small dispersion, showing that all trajectories are freezed, at least up to timescales of the order of $t_{obs}$. However, there is an intermediate coupling regime $0.3\lesssim U/J\lesssim 0.6$, that corresponds to the intermediate regime described when analyzing the relaxation of the kinetic energy in the previous section, for which the dispersion almost doest not changes. This confirms that inactive, non-ergodic classical trajectories are related with the existence of non-thermal metastable states in the quantum dynamics. Moreover, since the shift of the center of the histograms is continuous we may expect that for intermediate couplings there should be a coexistence of active and inactive trajectories, in which case the mobility of the trajectories would depend on the details of the initial conditions. This is indeed the case. This observation can be cast in the language of a dynamical phase transition~\cite{garrahan07_dyn_pt_kcm,garrahan09_dyn_pt_long,hedges2009dynamic,chandler10_dyn_pt_review,garrahan10_dyn_pt_quantum} taking place in the ensemble of classical trajectories.  To show this we first construct a canonical probability distribution in such ensemble, coupling the extensive-in-time order parameter $K\left[\psi(t)\right]$ with an intensive field $s$. This distribution will be proportional to
\bea
P_s\left[\psi(t)\right]\propto P_0\left[\psi(t)\right]e^{-s K\left[\psi(t)\right]}\,,
\eea
 where the $P_0\left[\psi(t)\right]$ is the $s=0$ probability distribution that we take as the Wigner function of the coherent state $p_C(\psi_0,\psi_0^*)$ as discussed earlier.%, in particular the $p_\delta$ since it will capture all the physics we are looking for having a quick convergence when doing a Monte-Carlo integration.

We will compute expectation values in the ensemble of trajectories by summing over all trajectories weighted with $P_s\left[\psi(t)\right]$ in the following fashion
\bea
\Omega_s=\left< \Omega\left[\psi(t)\right] \right>_s=\frac1{Z_s}\sum_{\psi(t)}P_s\left[\psi(t)\right] \Omega\left[\psi(t)\right]\,,
\eea
where $\Omega\left[\psi(t)\right]$ is a trajectory functional and $Z_s$ is the partition function
\bea
Z_s=\sum_{\psi(t)}P_s\left[\psi(t)\right]\,.
\eea
This way of averaging will give a different weight to mobile and immobile trajectories depending on the value of $s$: for larger $s$ more relative weight is assigned to inactive trajectories and viceversa.

\begin{figure}[htp]
\begin{center}
\includegraphics[width=\figwidth]{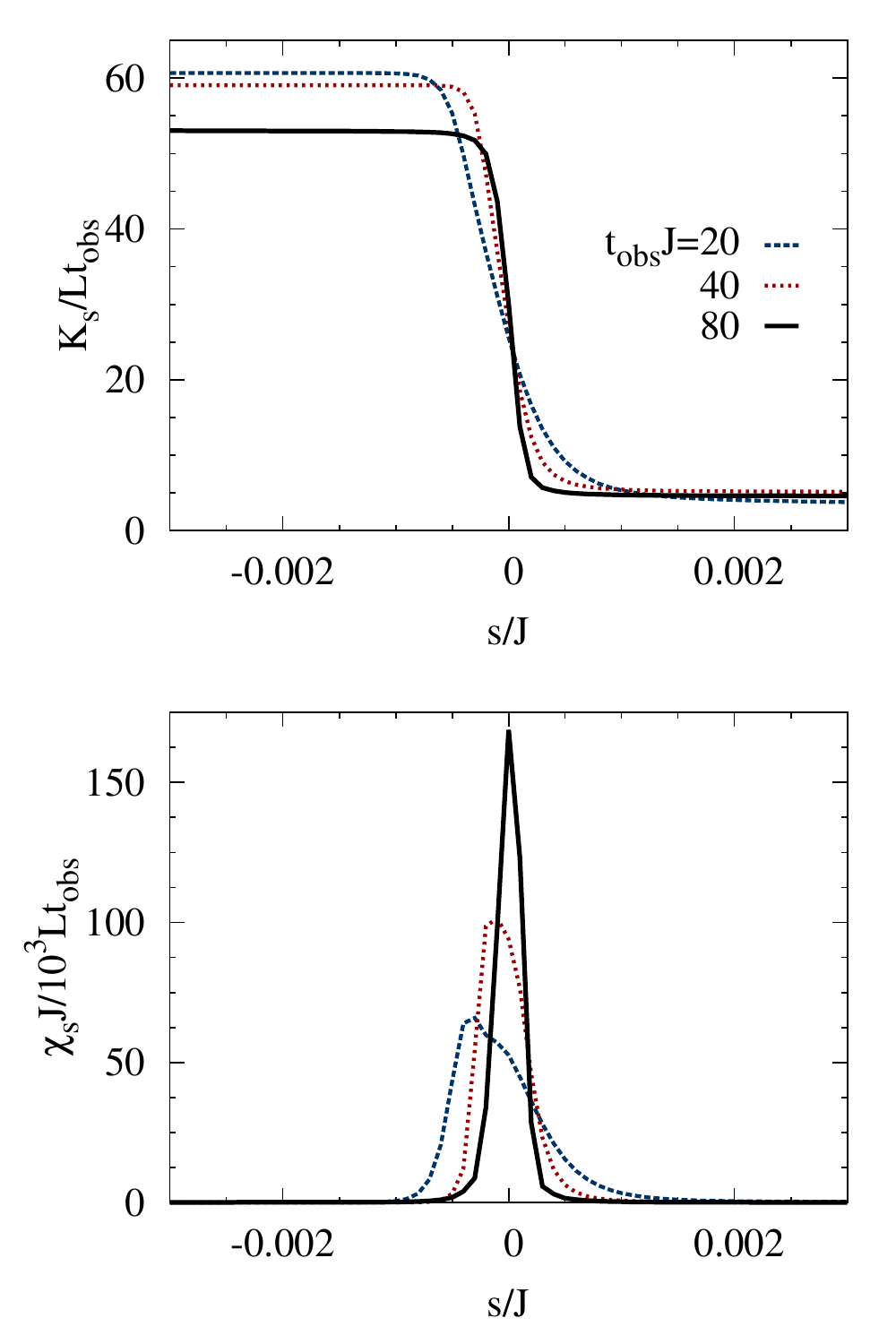}
\caption{\label{ps} Dynamical phase transition in trajectory space. Upper panel: Averaged order parameter $K_s$ as a function of the intensive field $s$ for an ensemble of $10^3$ trajectories sampled according to the coherent state Wigner function for $U/J=0.53$. Lower panel: Dynamical susceptibility $\chi_s$. }\label{fig:dyn_pt}
\end{center}
\end{figure}

In Fig.~\ref{fig:dyn_pt} we show the average order parameter $K_s$ as a function of $s$ for $U/J=0.53$, a value of $U/J$ for which we have observed coexistence of active and inactive trajectories. The behavior of $K_s$ mimics the behavior of the order parameter in a finite volume equilibrium phase transition: it shows a marked step between two well defined values corresponding to the two different phases in the ensemble (active and inactive phases) and, moreover, while increasing $t_{obs}$ (the analogous of the volume in equilibrium) the step in $K_s$ becomes more and more sharp. This can be appreciated more clearly looking at the susceptibility:
\be
\chi_s=-\frac{\partial K_s}{\partial s}=\langle (K[\psi(t)]-K_s)^2 \rangle_s,
\ee
which exhibits a peak that grows while increasing $t_{obs}$. A similar scaling behavior can be observed with increasing $L$. The critical value is located around zero, $s^*=0$. This dynamical phase transition is very similar to that found first in kinetically constrained models of glasses~\cite{garrahan07_dyn_pt_kcm} and then in atomistic models of glass formers~\cite{hedges2009dynamic} and quantum systems~\cite{garrahan10_dyn_pt_quantum}. The main difference that we observe with respect to the works~\cite{garrahan07_dyn_pt_kcm,hedges2009dynamic,garrahan10_dyn_pt_quantum} is that the present dynamic phase transition does not seems to be of first order. In a first order (dynamic or static) phase transition the distribution of the order parameter at coexistence is bimodal due to surface tension effects between the coexisting domains of different phases. In the case analyzed in this work the order parameter distribution at coexistence ($s\simeq0$ for $U/J=0.5$) is unimodal, as can be inferred looking at Fig.~\ref{fig:histo}, which is compatible with a continuous phase transition.

\begin{figure}[htp]
\begin{center}
\includegraphics[width=\largefigwidth]{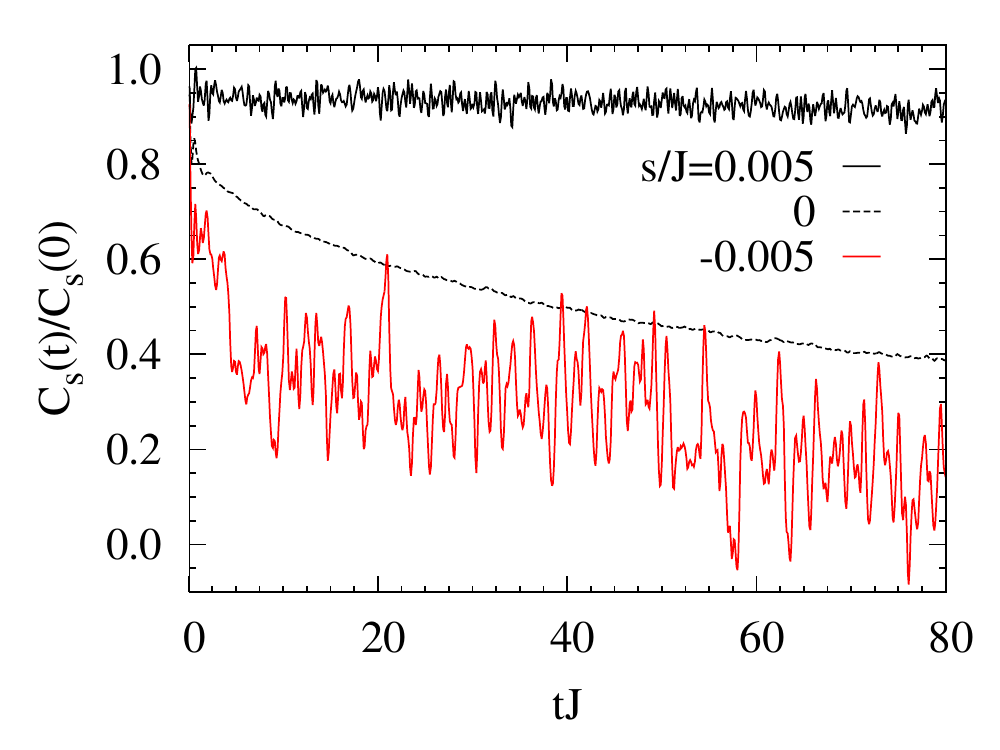}
\caption{\label{ps} Correlation function $C_s(t)$ as a function of time for $U/J=0.53$ in the active ($s<0$) and inactive ($s>0$) phases. We also show, for comparison, the correlation function at coexistence ($s=0$).}\label{fig:corr}
\end{center}
\end{figure}

To better characterize the physical properties of the phases we will discuss an ergodicity measure. We define the overlap correlation function,
\bea
C_s(t)=\sum_{j=1}^{L}\langle(n_j(t)-N)(n_j(0)-N)\rangle_s.
\eea
$C_s(t)$ quantifies the overlap between the configuration at time $t$ and the initial density configuration. The extent to which it is non-zero in the limit of large $t$ is a measure of nonergodicity. In Fig.~\ref{fig:corr} we show the correlation function $C_s(t)$ for the active (ergodic) and inactive (nonergodic) phases. In the active phase ($s<0$) the correlation function rapidly relaxes to a small value, while for the inactive phase ($s>0$) trajectories remain trapped in a single state throughout the observation time.

\begin{figure}[htp]
\begin{center}
\includegraphics[width=\largefigwidth]{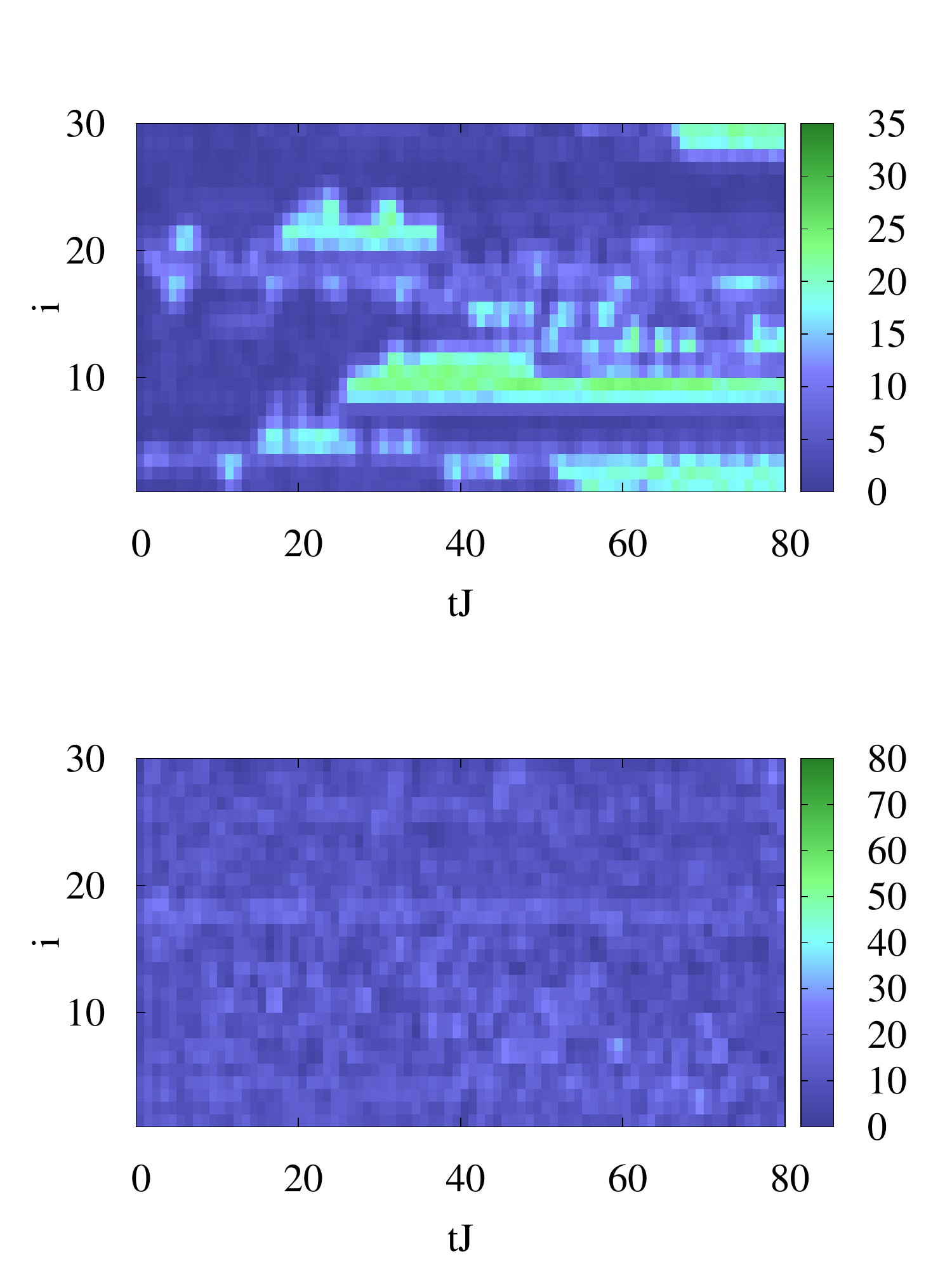}
\caption{\label{ps} False color plot of the local mobility $Q_{i}(t)$ for two representative trajectories corresponding to $U/J=0.6$ (upper panel) and $U/J=0.1$ (lower panel).}\label{fig:mob}
\end{center}
\end{figure}

In classical models of glasses such dynamical phase transition represents one of the main theories to account for an striking and distinctive feature of glassy materials: dynamic heterogeneity~\cite{chandler10_dyn_pt_review}. In contrast with normal fluids, the relaxation of glasses is heterogeneous in space: fast and slow regions are clustered together. We found that individual mean field trajectories (that correspond to single solutions of the Gross-Pitaevskii equation) exhibit the phenomenon of dynamical heterogeneity. We shall define a local measure of mobility $Q_{i}(t)$ as
\be
Q_i(t)=\vert n_i(t)-n_i(0)\vert,
\ee
that measures how much the in site density differs from the initial value. In Fig.~\ref{fig:mob} we show density plots of $Q_i(t)$ for representative trajectories corresponding to $U/J=0.1$ and $0.6$. For $U/J=0.1$ (the weak coupling regime) the dynamics of the system, as captured by the indicator $Q_i(t)$, is fairly homogeneous. For $U/J=0.6$ (entering the strong coupling regime), sites with high local mobility and low local mobility are clustered together in space-time. This is the signature of dynamic heterogeneity. It is remarkable that the simple Gross-Pitaevskii equation is able to generate such complicated dynamics.

The relation between the dynamical phase transition and the dynamical heterogeneity can be explained using a simple reasoning: the dynamical phase transition picture implies that for intermediate values of $U/J$ different initial conditions may trigger slow or fast dynamics, depending only on the details of the initial condition; if we consider a sufficiently large system with short range interactions, distant regions in space will not be correlated and, if one region has a ``fast'' initial condition and other a ``slow'' initial condition, dynamical heterogeneity may arise. Moreover, by the same argument, during the dynamical process slow regions may turn into fast regions and viceversa. On contrary to what is observed in \cite{nussinov2012dynamical,nussinov2012viewpoint}, the quantum dynamics does not show dynamical heterogeneity. The average needed to obtain the quantum dynamics from the individual classical trajectories washes out any initial condition dependent dynamical heterogeneous pattern, which is associated with the fact that the density profile has a well defined relaxation timescale as seen before.

\section{Conclusions}
\label{conc}

We have studied the dynamics of the Bose-Hubbard model following a quantum quench from an inhomogeneous initial state. Using the TWA we have analyzed the dynamics of the density profile of the system, that exhibits long-lived prethermalization plateaus for moderate and large couplings. We were able to relate this fact with the non-ergodic propoerties of the associated mean-field system. In particular, we have shown that the ensamble of mean field trajectories that are relevant to the quantum motion undergoes a dynamic phase transition from an active to an inactive phase, much alike to what is found in classical models of glasses. We have also shown that this transition is related to the phenomenon of dynamic heterogeneity in the classical trajectories, one of the hallmarks of glassiness. In sum, we believe that our work opens a new perspective on the dynamics of closed quantum systems by relating the physics of classical glasses to the well known phenomenon of prethermalization.

\section*{Acknowledgements}

We acknowledge very useful discussions with Andrea Gambassi and Pablo R. Ponte. ISL thanks ICTP where part of this project was done.
This work was partially supported by CONICET (PIP 0662),
ANPCyT (PICT 2010-1907) and UNLP (PID X497), Argentina.
%

%\bibliography{phys}

\begin{thebibliography}{41}
\expandafter\ifx\csname natexlab\endcsname\relax\def\natexlab#1{#1}\fi
\expandafter\ifx\csname bibnamefont\endcsname\relax
  \def\bibnamefont#1{#1}\fi
\expandafter\ifx\csname bibfnamefont\endcsname\relax
  \def\bibfnamefont#1{#1}\fi
\expandafter\ifx\csname citenamefont\endcsname\relax
  \def\citenamefont#1{#1}\fi
\expandafter\ifx\csname url\endcsname\relax
  \def\url#1{\texttt{#1}}\fi
\expandafter\ifx\csname urlprefix\endcsname\relax\def\urlprefix{URL }\fi
\providecommand{\bibinfo}[2]{#2}
\providecommand{\eprint}[2][]{\url{#2}}

\bibitem[{\citenamefont{Trotzky et~al.}(2012)\citenamefont{Trotzky, Chen,
  Flesch, McCulloch, SchollwÃ¶ck, Eisert, and
  Bloch}}]{trotzky11_relaxation_BH_10101010}
\bibinfo{author}{\bibfnamefont{S.}~\bibnamefont{Trotzky}},
  \bibinfo{author}{\bibfnamefont{Y.-A.} \bibnamefont{Chen}},
  \bibinfo{author}{\bibfnamefont{A.}~\bibnamefont{Flesch}},
  \bibinfo{author}{\bibfnamefont{I.~P.} \bibnamefont{McCulloch}},
  \bibinfo{author}{\bibfnamefont{U.}~\bibnamefont{SchollwÃ¶ck}},
  \bibinfo{author}{\bibfnamefont{J.}~\bibnamefont{Eisert}}, \bibnamefont{and}
  \bibinfo{author}{\bibfnamefont{I.}~\bibnamefont{Bloch}},
  \bibinfo{journal}{Nat. Phys.} \textbf{\bibinfo{volume}{8}},
  \bibinfo{pages}{325} (\bibinfo{year}{2012}).

\bibitem[{\citenamefont{Sorg et~al.}(2014)\citenamefont{Sorg, Vidmar, Pollet,
  and Heidrich-Meisner}}]{sorg14_thermalization_bose-hub}
\bibinfo{author}{\bibfnamefont{S.}~\bibnamefont{Sorg}},
  \bibinfo{author}{\bibfnamefont{L.}~\bibnamefont{Vidmar}},
  \bibinfo{author}{\bibfnamefont{L.}~\bibnamefont{Pollet}}, \bibnamefont{and}
  \bibinfo{author}{\bibfnamefont{F.}~\bibnamefont{Heidrich-Meisner}},
  \bibinfo{journal}{Phys. Rev. A} \textbf{\bibinfo{volume}{90}},
  \bibinfo{pages}{033606} (\bibinfo{year}{2014}).

\bibitem[{\citenamefont{Rigol et~al.}(2008)\citenamefont{Rigol, Dunjko, and
  Olshanii}}]{rigol08_mechanism_thermalization}
\bibinfo{author}{\bibfnamefont{M.}~\bibnamefont{Rigol}},
  \bibinfo{author}{\bibfnamefont{V.}~\bibnamefont{Dunjko}}, \bibnamefont{and}
  \bibinfo{author}{\bibfnamefont{M.}~\bibnamefont{Olshanii}},
  \bibinfo{journal}{Nature (London)} \textbf{\bibinfo{volume}{452}},
  \bibinfo{pages}{854} (\bibinfo{year}{2008}).

\bibitem[{\citenamefont{Rigol}(2014)}]{rigol14_quenches_thermo_limit}
\bibinfo{author}{\bibfnamefont{M.}~\bibnamefont{Rigol}},
  \bibinfo{journal}{Phys. Rev. Lett.} \textbf{\bibinfo{volume}{112}},
  \bibinfo{pages}{170601} (\bibinfo{year}{2014}).

\bibitem[{\citenamefont{Kinoshita et~al.}(2006)\citenamefont{Kinoshita, Wenger,
  and Weiss}}]{kinoshita06_non_thermalization}
\bibinfo{author}{\bibfnamefont{T.}~\bibnamefont{Kinoshita}},
  \bibinfo{author}{\bibfnamefont{T.}~\bibnamefont{Wenger}}, \bibnamefont{and}
  \bibinfo{author}{\bibfnamefont{D.~S.} \bibnamefont{Weiss}},
  \bibinfo{journal}{Nature (London)} \textbf{\bibinfo{volume}{440}},
  \bibinfo{pages}{900} (\bibinfo{year}{2006}).

\bibitem[{\citenamefont{Berges et~al.}(2004)\citenamefont{Berges, Bors\'anyi,
  and Wetterich}}]{berges04_prethermalization_idea}
\bibinfo{author}{\bibfnamefont{J.}~\bibnamefont{Berges}},
  \bibinfo{author}{\bibfnamefont{S.}~\bibnamefont{Bors\'anyi}},
  \bibnamefont{and}
  \bibinfo{author}{\bibfnamefont{C.}~\bibnamefont{Wetterich}},
  \bibinfo{journal}{Phys. Rev. Lett.} \textbf{\bibinfo{volume}{93}},
  \bibinfo{pages}{142002} (\bibinfo{year}{2004}).

\bibitem[{\citenamefont{Moeckel and
  Kehrein}(2008)}]{moeckel08_quench_hubbard_high_d}
\bibinfo{author}{\bibfnamefont{M.}~\bibnamefont{Moeckel}} \bibnamefont{and}
  \bibinfo{author}{\bibfnamefont{S.}~\bibnamefont{Kehrein}},
  \bibinfo{journal}{Phys. Rev. Lett.} \textbf{\bibinfo{volume}{100}},
  \bibinfo{pages}{175702} (\bibinfo{year}{2008}).

\bibitem[{\citenamefont{Eckstein et~al.}(2009)\citenamefont{Eckstein, Kollar,
  and Werner}}]{eckstein09_thermalization_quench_fermi_hubbard}
\bibinfo{author}{\bibfnamefont{M.}~\bibnamefont{Eckstein}},
  \bibinfo{author}{\bibfnamefont{M.}~\bibnamefont{Kollar}}, \bibnamefont{and}
  \bibinfo{author}{\bibfnamefont{P.}~\bibnamefont{Werner}},
  \bibinfo{journal}{Phys. Rev. Lett.} \textbf{\bibinfo{volume}{103}},
  \bibinfo{pages}{056403} (\bibinfo{year}{2009}).

\bibitem[{\citenamefont{Kollar et~al.}(2011)\citenamefont{Kollar, Wolf, and
  Eckstein}}]{kollar11_gge_pretherm}
\bibinfo{author}{\bibfnamefont{M.}~\bibnamefont{Kollar}},
  \bibinfo{author}{\bibfnamefont{F.~A.} \bibnamefont{Wolf}}, \bibnamefont{and}
  \bibinfo{author}{\bibfnamefont{M.}~\bibnamefont{Eckstein}},
  \bibinfo{journal}{Phys. Rev. B} \textbf{\bibinfo{volume}{84}},
  \bibinfo{pages}{054304} (\bibinfo{year}{2011}).

\bibitem[{\citenamefont{Nessi et~al.}(2014)\citenamefont{Nessi, Iucci, and
  Cazalilla}}]{nessi14_gge_fermi_liquid}
\bibinfo{author}{\bibfnamefont{N.}~\bibnamefont{Nessi}},
  \bibinfo{author}{\bibfnamefont{A.}~\bibnamefont{Iucci}}, \bibnamefont{and}
  \bibinfo{author}{\bibfnamefont{M.~A.} \bibnamefont{Cazalilla}},
  \bibinfo{journal}{Phys. Rev. Lett.} \textbf{\bibinfo{volume}{113}},
  \bibinfo{pages}{210402} (\bibinfo{year}{2014}).

\bibitem[{\citenamefont{Gring et~al.}(2012)\citenamefont{Gring, Kuhnert,
  Langen, Kitagawa, Rauer, Schreitl, Mazets, Smith, Demler, and
  Schmiedmayer}}]{gring12_prethermalization_isolated_bose_gas}
\bibinfo{author}{\bibfnamefont{M.}~\bibnamefont{Gring}},
  \bibinfo{author}{\bibfnamefont{M.}~\bibnamefont{Kuhnert}},
  \bibinfo{author}{\bibfnamefont{T.}~\bibnamefont{Langen}},
  \bibinfo{author}{\bibfnamefont{T.}~\bibnamefont{Kitagawa}},
  \bibinfo{author}{\bibfnamefont{B.}~\bibnamefont{Rauer}},
  \bibinfo{author}{\bibfnamefont{M.}~\bibnamefont{Schreitl}},
  \bibinfo{author}{\bibfnamefont{I.}~\bibnamefont{Mazets}},
  \bibinfo{author}{\bibfnamefont{D.~A.} \bibnamefont{Smith}},
  \bibinfo{author}{\bibfnamefont{E.}~\bibnamefont{Demler}}, \bibnamefont{and}
  \bibinfo{author}{\bibfnamefont{J.}~\bibnamefont{Schmiedmayer}},
  \bibinfo{journal}{Science} \textbf{\bibinfo{volume}{337}},
  \bibinfo{pages}{1318} (\bibinfo{year}{2012}).

\bibitem[{\citenamefont{Essler et~al.}(2014)\citenamefont{Essler, Kehrein,
  Manmana, and Robinson}}]{essler13_quench_tuneable_integrability_breaking}
\bibinfo{author}{\bibfnamefont{F.~H.~L.} \bibnamefont{Essler}},
  \bibinfo{author}{\bibfnamefont{S.}~\bibnamefont{Kehrein}},
  \bibinfo{author}{\bibfnamefont{S.~R.} \bibnamefont{Manmana}},
  \bibnamefont{and} \bibinfo{author}{\bibfnamefont{N.~J.}
  \bibnamefont{Robinson}}, \bibinfo{journal}{Phys. Rev. B}
  \textbf{\bibinfo{volume}{89}}, \bibinfo{pages}{165104}
  (\bibinfo{year}{2014}).

\bibitem[{\citenamefont{Marcuzzi et~al.}(2013)\citenamefont{Marcuzzi, Marino,
  Gambassi, and Silva}}]{marcuzzi13_pretherm_ising}
\bibinfo{author}{\bibfnamefont{M.}~\bibnamefont{Marcuzzi}},
  \bibinfo{author}{\bibfnamefont{J.}~\bibnamefont{Marino}},
  \bibinfo{author}{\bibfnamefont{A.}~\bibnamefont{Gambassi}}, \bibnamefont{and}
  \bibinfo{author}{\bibfnamefont{A.}~\bibnamefont{Silva}},
  \bibinfo{journal}{Phys. Rev. Lett.} \textbf{\bibinfo{volume}{111}},
  \bibinfo{pages}{197203} (\bibinfo{year}{2013}).

\bibitem[{\citenamefont{Stark and Kollar}(2013)}]{stark13_kinetic_description}
\bibinfo{author}{\bibfnamefont{M.}~\bibnamefont{Stark}} \bibnamefont{and}
  \bibinfo{author}{\bibfnamefont{M.}~\bibnamefont{Kollar}},
  \bibinfo{journal}{unpublished}  (\bibinfo{year}{2013}),
  \eprint{arXiv:1308.1610}.

\bibitem[{\citenamefont{Langen et~al.}(2015)\citenamefont{Langen, Erne, Geiger,
  Rauer, Schweigler, Kuhnert, Rohringer, Mazets, Gasenzer, and
  Schmiedmayer}}]{langen15_quench_lieb-liniger}
\bibinfo{author}{\bibfnamefont{T.}~\bibnamefont{Langen}},
  \bibinfo{author}{\bibfnamefont{S.}~\bibnamefont{Erne}},
  \bibinfo{author}{\bibfnamefont{R.}~\bibnamefont{Geiger}},
  \bibinfo{author}{\bibfnamefont{B.}~\bibnamefont{Rauer}},
  \bibinfo{author}{\bibfnamefont{T.}~\bibnamefont{Schweigler}},
  \bibinfo{author}{\bibfnamefont{M.}~\bibnamefont{Kuhnert}},
  \bibinfo{author}{\bibfnamefont{W.}~\bibnamefont{Rohringer}},
  \bibinfo{author}{\bibfnamefont{I.~E.} \bibnamefont{Mazets}},
  \bibinfo{author}{\bibfnamefont{T.}~\bibnamefont{Gasenzer}}, \bibnamefont{and}
  \bibinfo{author}{\bibfnamefont{J.}~\bibnamefont{Schmiedmayer}},
  \bibinfo{journal}{unpublished}  (\bibinfo{year}{2015}),
  \eprint{arXiv:1411.7185}.

\bibitem[{\citenamefont{Kollath et~al.}(2007)\citenamefont{Kollath, L\"auchli,
  and Altman}}]{kollath07_quench_BH}
\bibinfo{author}{\bibfnamefont{C.}~\bibnamefont{Kollath}},
  \bibinfo{author}{\bibfnamefont{A.~M.} \bibnamefont{L\"auchli}},
  \bibnamefont{and} \bibinfo{author}{\bibfnamefont{E.}~\bibnamefont{Altman}},
  \bibinfo{journal}{Phys. Rev. Lett.} \textbf{\bibinfo{volume}{98}},
  \bibinfo{pages}{180601} (\bibinfo{year}{2007}).

\bibitem[{\citenamefont{Roux}(2009)}]{roux09_quench_bose-hub}
\bibinfo{author}{\bibfnamefont{G.}~\bibnamefont{Roux}}, \bibinfo{journal}{Phys.
  Rev. A} \textbf{\bibinfo{volume}{79}}, \bibinfo{pages}{021608}
  (\bibinfo{year}{2009}).

\bibitem[{\citenamefont{Carleo et~al.}(2012{\natexlab{a}})\citenamefont{Carleo,
  Becca, Schir{\'o}, and Fabrizio}}]{carleo2012localization}
\bibinfo{author}{\bibfnamefont{G.}~\bibnamefont{Carleo}},
  \bibinfo{author}{\bibfnamefont{F.}~\bibnamefont{Becca}},
  \bibinfo{author}{\bibfnamefont{M.}~\bibnamefont{Schir{\'o}}},
  \bibnamefont{and} \bibinfo{author}{\bibfnamefont{M.}~\bibnamefont{Fabrizio}},
  \bibinfo{journal}{Scientific Reports} \textbf{\bibinfo{volume}{2}}
  (\bibinfo{year}{2012}{\natexlab{a}}).

\bibitem[{\citenamefont{Bloch et~al.}(2008)\citenamefont{Bloch, Dalibard, and
  Zwerger}}]{bloch08_cold_atoms_optical_lattices_review}
\bibinfo{author}{\bibfnamefont{I.}~\bibnamefont{Bloch}},
  \bibinfo{author}{\bibfnamefont{J.}~\bibnamefont{Dalibard}}, \bibnamefont{and}
  \bibinfo{author}{\bibfnamefont{W.}~\bibnamefont{Zwerger}},
  \bibinfo{journal}{Rev. Mod. Phys.} \textbf{\bibinfo{volume}{80}},
  \bibinfo{pages}{885} (\bibinfo{year}{2008}).

\bibitem[{\citenamefont{Natu et~al.}(2011)\citenamefont{Natu, Hazzard, and
  Mueller}}]{natu11_local_eq_inhom_BEC}
\bibinfo{author}{\bibfnamefont{S.~S.} \bibnamefont{Natu}},
  \bibinfo{author}{\bibfnamefont{K.~R.~A.} \bibnamefont{Hazzard}},
  \bibnamefont{and} \bibinfo{author}{\bibfnamefont{E.~J.}
  \bibnamefont{Mueller}}, \bibinfo{journal}{Phys. Rev. Lett.}
  \textbf{\bibinfo{volume}{106}}, \bibinfo{pages}{125301}
  (\bibinfo{year}{2011}).

\bibitem[{\citenamefont{Hung et~al.}(2010)\citenamefont{Hung, Zhang, Gemelke,
  and Chin}}]{hung10_slow_transport_inhom_BEC}
\bibinfo{author}{\bibfnamefont{C.-L.} \bibnamefont{Hung}},
  \bibinfo{author}{\bibfnamefont{X.}~\bibnamefont{Zhang}},
  \bibinfo{author}{\bibfnamefont{N.}~\bibnamefont{Gemelke}}, \bibnamefont{and}
  \bibinfo{author}{\bibfnamefont{C.}~\bibnamefont{Chin}},
  \bibinfo{journal}{Phys. Rev. Lett.} \textbf{\bibinfo{volume}{104}},
  \bibinfo{pages}{160403} (\bibinfo{year}{2010}).

\bibitem[{\citenamefont{McKay et~al.}(2015)\citenamefont{McKay, Ray, Natu,
  Russ, Ceperley, , and DeMarco}}]{mckay15_metastability_inhom_BEC}
\bibinfo{author}{\bibfnamefont{D.}~\bibnamefont{McKay}},
  \bibinfo{author}{\bibfnamefont{U.}~\bibnamefont{Ray}},
  \bibinfo{author}{\bibfnamefont{S.}~\bibnamefont{Natu}},
  \bibinfo{author}{\bibfnamefont{P.}~\bibnamefont{Russ}},
  \bibinfo{author}{\bibfnamefont{D.}~\bibnamefont{Ceperley}}, ,
  \bibnamefont{and} \bibinfo{author}{\bibfnamefont{B.}~\bibnamefont{DeMarco}},
  \bibinfo{journal}{Phys. Rev. Lett.} \textbf{\bibinfo{volume}{91}},
  \bibinfo{pages}{023625} (\bibinfo{year}{2015}).

\bibitem[{\citenamefont{Srednicki}(1994)}]{srednicki94_ETH}
\bibinfo{author}{\bibfnamefont{M.}~\bibnamefont{Srednicki}},
  \bibinfo{journal}{Phys. Rev. E} \textbf{\bibinfo{volume}{50}},
  \bibinfo{pages}{888} (\bibinfo{year}{1994}).

\bibitem[{\citenamefont{Cosme and Fialko}(2014)}]{cosme2014thermalization}
\bibinfo{author}{\bibfnamefont{J.~G.} \bibnamefont{Cosme}} \bibnamefont{and}
  \bibinfo{author}{\bibfnamefont{O.}~\bibnamefont{Fialko}},
  \bibinfo{journal}{Phys. Rev. A} \textbf{\bibinfo{volume}{90}},
  \bibinfo{pages}{053602} (\bibinfo{year}{2014}).

\bibitem[{\citenamefont{Polkovnikov}(2010)}]{polkovnikov2010phase}
\bibinfo{author}{\bibfnamefont{A.}~\bibnamefont{Polkovnikov}},
  \bibinfo{journal}{Annals of Physics} \textbf{\bibinfo{volume}{325}},
  \bibinfo{pages}{1790} (\bibinfo{year}{2010}).

\bibitem[{\citenamefont{Jona-Lasinio and Presilla}(1996)}]{jona1996chaotic}
\bibinfo{author}{\bibfnamefont{G.}~\bibnamefont{Jona-Lasinio}}
  \bibnamefont{and} \bibinfo{author}{\bibfnamefont{C.}~\bibnamefont{Presilla}},
  \bibinfo{journal}{Phys. Rev. Lett.} \textbf{\bibinfo{volume}{77}},
  \bibinfo{pages}{4322} (\bibinfo{year}{1996}).

\bibitem[{\citenamefont{Cassidy et~al.}(2009)\citenamefont{Cassidy, Mason,
  Dunjko, and Olshanii}}]{cassidy09_quench_bose-hub}
\bibinfo{author}{\bibfnamefont{A.}~\bibnamefont{Cassidy}},
  \bibinfo{author}{\bibfnamefont{D.}~\bibnamefont{Mason}},
  \bibinfo{author}{\bibfnamefont{V.}~\bibnamefont{Dunjko}}, \bibnamefont{and}
  \bibinfo{author}{\bibfnamefont{M.}~\bibnamefont{Olshanii}},
  \bibinfo{journal}{Phys. Rev. Lett.} \textbf{\bibinfo{volume}{105}},
  \bibinfo{pages}{025302} (\bibinfo{year}{2009}).

\bibitem[{\citenamefont{Castellani and
  Cavagna}(2005)}]{castellani05_spin-glass_pedestrians}
\bibinfo{author}{\bibfnamefont{T.}~\bibnamefont{Castellani}} \bibnamefont{and}
  \bibinfo{author}{\bibfnamefont{A.}~\bibnamefont{Cavagna}},
  \bibinfo{journal}{J. Stat. Mech.: Theor. Exp.}  (\bibinfo{year}{2005}).

\bibitem[{\citenamefont{Cavagna}(2009)}]{cavagna09_supercooled_pedestrians}
\bibinfo{author}{\bibfnamefont{A.}~\bibnamefont{Cavagna}},
  \bibinfo{journal}{Phys. Rep.} \textbf{\bibinfo{volume}{476}},
  \bibinfo{pages}{51} (\bibinfo{year}{2009}).

\bibitem[{\citenamefont{Chandler and
  Garrahan}(2010)}]{chandler10_dyn_pt_review}
\bibinfo{author}{\bibfnamefont{D.}~\bibnamefont{Chandler}} \bibnamefont{and}
  \bibinfo{author}{\bibfnamefont{J.~P.} \bibnamefont{Garrahan}},
  \bibinfo{journal}{Annu. Rev. Phys. Chem.} \textbf{\bibinfo{volume}{61}},
  \bibinfo{pages}{191} (\bibinfo{year}{2010}).

\bibitem[{\citenamefont{Berthier et~al.}(2011)\citenamefont{Berthier, Biroli,
  Bouchaud, Cipelletti, and van Sarloos}}]{berthier11_dyn_heterogeneity_kcm}
\bibinfo{author}{\bibfnamefont{L.}~\bibnamefont{Berthier}},
  \bibinfo{author}{\bibfnamefont{G.}~\bibnamefont{Biroli}},
  \bibinfo{author}{\bibfnamefont{J.-P.} \bibnamefont{Bouchaud}},
  \bibinfo{author}{\bibfnamefont{L.}~\bibnamefont{Cipelletti}},
  \bibnamefont{and} \bibinfo{author}{\bibfnamefont{W.}~\bibnamefont{van
  Sarloos}}, \emph{\bibinfo{title}{Dynamical Heterogeneities in Glasses,
  Colloids and Granular Media}} (\bibinfo{publisher}{Oxford Science
  Publications}, \bibinfo{year}{2011}).

\bibitem[{\citenamefont{Garrahan et~al.}(2007)\citenamefont{Garrahan, Jack,
  Lecomte, Pitard, van Duijvendijk, and van Wijland}}]{garrahan07_dyn_pt_kcm}
\bibinfo{author}{\bibfnamefont{J.~P.} \bibnamefont{Garrahan}},
  \bibinfo{author}{\bibfnamefont{R.~L.} \bibnamefont{Jack}},
  \bibinfo{author}{\bibfnamefont{V.}~\bibnamefont{Lecomte}},
  \bibinfo{author}{\bibfnamefont{E.}~\bibnamefont{Pitard}},
  \bibinfo{author}{\bibfnamefont{K.}~\bibnamefont{van Duijvendijk}},
  \bibnamefont{and} \bibinfo{author}{\bibfnamefont{F.}~\bibnamefont{van
  Wijland}}, \bibinfo{journal}{Phys. Rev. Lett.} \textbf{\bibinfo{volume}{98}},
  \bibinfo{pages}{195702} (\bibinfo{year}{2007}).

\bibitem[{\citenamefont{Garrahan et~al.}(2009)\citenamefont{Garrahan, Jack,
  Lecomte, Pitard, van Duijvendijk, and van Wijland}}]{garrahan09_dyn_pt_long}
\bibinfo{author}{\bibfnamefont{J.~P.} \bibnamefont{Garrahan}},
  \bibinfo{author}{\bibfnamefont{R.~L.} \bibnamefont{Jack}},
  \bibinfo{author}{\bibfnamefont{V.}~\bibnamefont{Lecomte}},
  \bibinfo{author}{\bibfnamefont{E.}~\bibnamefont{Pitard}},
  \bibinfo{author}{\bibfnamefont{K.}~\bibnamefont{van Duijvendijk}},
  \bibnamefont{and} \bibinfo{author}{\bibfnamefont{F.}~\bibnamefont{van
  Wijland}}, \bibinfo{journal}{J. Phys. A: Math. Theor.}
  \textbf{\bibinfo{volume}{42}}, \bibinfo{pages}{075007}
  (\bibinfo{year}{2009}).

\bibitem[{\citenamefont{Hedges et~al.}(2009)\citenamefont{Hedges, Jack,
  Garrahan, and Chandler}}]{hedges2009dynamic}
\bibinfo{author}{\bibfnamefont{L.~O.} \bibnamefont{Hedges}},
  \bibinfo{author}{\bibfnamefont{R.~L.} \bibnamefont{Jack}},
  \bibinfo{author}{\bibfnamefont{J.~P.} \bibnamefont{Garrahan}},
  \bibnamefont{and} \bibinfo{author}{\bibfnamefont{D.}~\bibnamefont{Chandler}},
  \bibinfo{journal}{Science} \textbf{\bibinfo{volume}{323}},
  \bibinfo{pages}{1309} (\bibinfo{year}{2009}).

\bibitem[{\citenamefont{Polkovnikov et~al.}(2002)\citenamefont{Polkovnikov,
  Sachdev, and Girvin}}]{polkovnikov2002evolution}
\bibinfo{author}{\bibfnamefont{A.}~\bibnamefont{Polkovnikov}},
  \bibinfo{author}{\bibfnamefont{S.}~\bibnamefont{Sachdev}}, \bibnamefont{and}
  \bibinfo{author}{\bibfnamefont{S.~M.} \bibnamefont{Girvin}},
  \bibinfo{journal}{Phys. Rev. A} \textbf{\bibinfo{volume}{66}},
  \bibinfo{pages}{053607} (\bibinfo{year}{2002}).

\bibitem[{\citenamefont{Olsen and Bradley}(2009)}]{olsen09_wigner_sampling}
\bibinfo{author}{\bibfnamefont{M.~K.} \bibnamefont{Olsen}} \bibnamefont{and}
  \bibinfo{author}{\bibfnamefont{A.~S.} \bibnamefont{Bradley}},
  \bibinfo{journal}{Optics Communications} \textbf{\bibinfo{volume}{282}},
  \bibinfo{pages}{3924} (\bibinfo{year}{2009}).

\bibitem[{\citenamefont{Polkovnikov}(2003)}]{polkovnikov2003evolution}
\bibinfo{author}{\bibfnamefont{A.}~\bibnamefont{Polkovnikov}},
  \bibinfo{journal}{Phys. Rev. A} \textbf{\bibinfo{volume}{68}},
  \bibinfo{pages}{033609} (\bibinfo{year}{2003}).

\bibitem[{\citenamefont{Carleo et~al.}(2012{\natexlab{b}})\citenamefont{Carleo,
  Becca, Schir\'{o}, and Fabrizio}}]{carleo12_glassy_bose-hubbard}
\bibinfo{author}{\bibfnamefont{G.}~\bibnamefont{Carleo}},
  \bibinfo{author}{\bibfnamefont{F.}~\bibnamefont{Becca}},
  \bibinfo{author}{\bibfnamefont{M.}~\bibnamefont{Schir\'{o}}},
  \bibnamefont{and} \bibinfo{author}{\bibfnamefont{M.}~\bibnamefont{Fabrizio}},
  \bibinfo{journal}{Scientific Reports} \textbf{\bibinfo{volume}{2}},
  \bibinfo{pages}{243} (\bibinfo{year}{2012}{\natexlab{b}}).

\bibitem[{\citenamefont{Garrahan and
  Lesanovsky}(2010)}]{garrahan10_dyn_pt_quantum}
\bibinfo{author}{\bibfnamefont{J.~P.} \bibnamefont{Garrahan}} \bibnamefont{and}
  \bibinfo{author}{\bibfnamefont{I.}~\bibnamefont{Lesanovsky}},
  \bibinfo{journal}{Phys. Rev. Lett.} \textbf{\bibinfo{volume}{104}},
  \bibinfo{pages}{160601} (\bibinfo{year}{2010}).

\bibitem[{\citenamefont{Nussinov et~al.}(2012)\citenamefont{Nussinov, Johnson,
  Graf, and Balatsky}}]{nussinov2012dynamical}
\bibinfo{author}{\bibfnamefont{Z.}~\bibnamefont{Nussinov}},
  \bibinfo{author}{\bibfnamefont{P.}~\bibnamefont{Johnson}},
  \bibinfo{author}{\bibfnamefont{M.~J.} \bibnamefont{Graf}}, \bibnamefont{and}
  \bibinfo{author}{\bibfnamefont{A.~V.} \bibnamefont{Balatsky}},
  \bibinfo{journal}{arXiv preprint arXiv:1209.3823}  (\bibinfo{year}{2012}).

\bibitem[{\citenamefont{Nussinov}(2012)}]{nussinov2012viewpoint}
\bibinfo{author}{\bibfnamefont{Z.}~\bibnamefont{Nussinov}},
  \bibinfo{journal}{arXiv preprint arXiv:1203.4648}  (\bibinfo{year}{2012}).

\end{thebibliography}

\end{document}